# THE X-RAY R AQUARII: A TWO-SIDED JET AND CENTRAL SOURCE

Running head: (x-ray R Aquarii: a two-sided jet and central source)

November 19, 2001

E. Kellogg[1], J. A. Pedelty[2], and R.G. Lyon[3]

[1]Harvard/Smithsonian Center for Astrophysics, 60 Garden St. MS-27 Cambridge, MA 02138

[2]Biospheric Sciences Branch, Code 923, NASA's Goddard Space Flight Center, Greenbelt, MD 20771

[3]Instrument Technology Center, Code 550, NASA's Goddard Space Flight Center, Greenbelt, MD 20771

## ABSTRACT

We report Chandra ACIS-S3 x-ray imaging and spectroscopy of the R Aquarii binary system that show a spatially resolved two-sided jet and an unresolved central source. This is the first published report of such an x-ray jet seen in an evolved stellar system comprised of ~2-3 $M_\odot$ (also see Kellogg et al 2000). At $E < 1$ keV, the x-ray jet extends both to the northeast and southwest relative to the central binary system. At $1 < E < 7.1$ keV, R Aqr is a point-like source centered on the star system. While both 3.5-cm radio continuum emission and x-ray emission appear coincident in projection and have maximum intensities at ~7.5" NE of the central binary system, the next strongest x-ray component is located ~30" SW of the central binary system and has no radio continuum counterpart. The x-ray jets are likely shock heated in the recent past, and are not in thermal equilibrium. The strongest SW x-ray jet component may have been shocked recently since there is no relic radio emission as expected from an older shock. At the position of the central binary, we detect x-ray emission below 1.6 keV consistent with blackbody emission at $T \sim 2 \times 10^6$ K. There is also a prominent 6.4 keV feature, a possible fluorescence or collisionally excited Fe K$\alpha$ line from an accretion disk or from the wind of the giant star. For this excitation to occur, there must be an unseen hard source of x-rays or particles in the immediate vicinity of the hot star. Such a source would be hidden from view by the surrounding edge-on accretion disk.

## 1. INTRODUCTION

R Aqr is a symbiotic stellar system: a mass losing ~1-2 $M_\odot$ Mira-like long period variable (LPV) with a 387 day period and a ~1 $M_\odot$ hot companion with many of its features explained by an accretion disk model. The companion is believed to give rise to the non-relativistic jet seen at UV, optical, and radio wavelengths. Since no UV continuum is observed, the accretion disk is believed to be edge-on (Kafatos, Michalitsianos, & Hollis 1986). The binary system orbit has been characterized as highly inclined to the line of sight ($i \sim 70^o$) with large eccentricity ($\varepsilon \sim 0.8$), a semi-major axis of $\sim 2.6 \times 10^{14}$ cm, an orbital period of ~44 yr, and at a distance of 200 pc (see Hollis, Pedelty, & Lyon 1997, and references therein). Hence, the northeast (NE)-southwest (SW) oriented jet is probably refueled episodically with increased activity at periastron since the hot companion passes through the LPV's outer envelope. It is further posited that the binary system was at apastron circa 1996 (Hollis et al 1997a), so neither star should influence the other for many years near this epoch.

From the early 1980s to the present epoch, the dominant NE radio jet has undergone NE motion outward from the central HII region and counterclockwise rotation i.e. precession about the central HII region (Hollis, Pedelty, & Kafatos 1997b). UV, optical, and radio observations of the NE jet are described as resulting from shock excitation (e.g., Hollis et al. 1991; Hollis, Dorband, & Yusef-Zadeh 1992; Hollis et al 1997b) as the jet motion impacts the considerable circumstellar material from LPV mass loss. For example, the thermal bremsstrahlung radiation seen in the radio regime is suggested as a relic of a passing shock front. Shock modeling of the R Aqr NE radio and UV jets suggests that temperatures should be on the order of that required for x-ray emission (Hollis et al 1997b). The R Aqr system was first reported as an unresolved Einstein x-ray source by Jura & Helfand (1984). It was later observed with EXOSAT (Viotti et al. 1987) and ROSAT (Hünsch et al 1998), so we planned a Chandra observation to determine the spatial and spectral structure and nature of the R Aqr x-ray emission.

## 2. OBSERVATIONS AND RESULTS

### 2.1. Chandra Imagery and Spectra

Chandra observed R Aqr for 24.5 ks on 2000 September 10 (JD 2451797.7 start time) with the Advanced CCD Imaging Spectrometer (ACIS)-S3 back-illuminated chip, which records photon energy for imaging and spectroscopy. The ACIS-S3 pixel size is 0.492" (CXC 2000). The Chandra ACIS-S3 on-axis point-spread function in the 0.3 to 1.0 keV range has 50% and 80% encircled energy radii of ~0.418" and ~0.685", respectively (CXC 2000). A spatial image was constructed from the standard pipeline processed data. X-ray events were extracted from a 0.38 arcmin$^2$ region (5668 pixels) encompassing the jets and central source, yielding 1540 total counts in 22,717 s, or 0.068 counts s$^{-1}$.

Figure 1 is a contour image of smoothed Chandra data showing a resolved NE jet, an unresolved central source, and a resolved multi-component SW jet. The Figure 1 image was shifted in position by ~0.4" in right ascension and ~0.6" in declination to align the peaks of the unresolved source in the x-ray and radio images (see §2.2 below). The Chandra absolute position uncertainty has a 1-$\sigma$ error circle radius of 0.6" (CXC 2000) while the radio positions are accurate to 0.1", one tenth of the restoring beam size.

Spectra of the NE jet, the central source, and the SW jet were extracted using CIAO Chandra analysis routines and XSPEC (Arnaud 1996). Figure 2 shows the extraction regions, spectra and fits. In spite of small numbers of counts, the spectra required complex models for reasonable fits. A nonequilibrium collisional excitation model was used for the jets (Borkowski et al 2001) with Anders & Grevesse (1989) (hereafter AG) initial abundances, with varying abundances for the SW jet (see Table 1). Data were binned into 15 counts $ch^{-1}$ for the NE and SW jets and 114 eV channels for the central region. We fitted the interstellar column density (the XSPEC WABS model). Results were consistent with radio observations of $1.84 \times 10^{20}$ $cm^{-2}$ (Stark et al 1992). The NE jet shows two peaks, at ~400 and 550 eV, and has few counts above 1 keV. The Raymond & Smith (1977) model gave a temperature in the $2 \times 10^6$ K range with <1% confidence. It gave only a peak at the O VII line complex, ~570 eV, increasing only to 6% confidence with variable abundances (the calculated N peak in that case was too broad to fit the data well, due to a complex of lines present in the model). Therefore, we used the NEI collisional excitation XSPEC model (Borkowski et al 2001). This gave an acceptable fit, including a second peak at ~430 eV. The SW jet spectrum has only one peak at ~550 eV. A Raymond-Smith fit gave a calcu-lated peak at a higher energy than observed and only 7% confidence, not improved by varying the abundances from the AG values. The XSPEC mekal models gave similar results. A good fit was obtained using the NEI model only with relative O abundances ≥20. At relative O abundances < 20, the calculated O peak was at a higher energy than the data peak. More complex collisional excitation models having more free parameters did not improve the fit.

In Figure 2, we also show a fit to a blackbody for the central source at low energies, and nine gaussians at expected K line energies of Fe and the lighter even Z elements, down to 1.6 keV, where the blackbody model dominates. We quote the blackbody fit parameters in Table 1, although we could not reject equilibrium optically thin hot gas models at the 50% confidence level. The hard central source shows a feature containing 16 counts with a flux of $4.9 \pm 1.7 \times 10^{-14}$ ergs cm$^{-2}$ s$^{-1}$ at $6.41 \pm 0.04$ keV. This could be Fe I (i.e., K$\alpha$), or up to Fe IV, based on the calculations tabulated in Kaastra and Mewe (1993), or up to Fe X based on the measurements of Decaux et al (1997). At very low statistical significance, there are also suggestions of emission from Fe K$\beta$ at 7.1 keV, 3 counts, and emission from K$\alpha$ lines of Ca, V, and Cr: 7 counts at 3.7, 6 counts at 5.0, and 4 counts at 5.4 keV, respectively. The expected number of background counts in each channel is a few $\times 10^{-2}$; there are essentially *no* background counts in this energy range. This was checked by displacing the central source region about 1″; it contained no counts. Our data are very likely real x-ray emission events from the central source.

We observed a flux of $1.0 \pm 0.2 \times 10^{-13}$ erg cm$^{-2}$ s$^{-1}$ at $E < 2$ keV from R Aqr. Hünsch et al (1998) observed 5× greater flux in that energy range during the ROSAT all sky survey, epoch 1990-1991. We also observed an additional $1.0 \pm 0.25 \times 10^{-13}$ erg cm$^{-2}$ s$^{-1}$ at $E > 2$ keV, not visible to ROSAT. Viotti et al (1987) inferred ~$2 \times 10^{-11}$ erg cm$^{-2}$ s$^{-1}$ by estimating the spectrum of the source. They also interpreted the Einstein report as an upper limit that corresponds to ~ $2.5 \times 10^{-13}$ erg cm$^{-2}$ s$^{-1}$. We conclude that R Aqr is several times fainter in our data than it was in the 1980s and early 1990s.

## 2.2. VLA Imagery

R Aqr was observed on 1999 November 1, 4, and 7 with the NRAO[1] Very Large Array in the B configuration. Two sets of receivers were combined to observe a 100 MHz band centered at a frequency of 8460.1 MHz (~3.5 cm wavelength). A total of ~2 million visibilities were accumulated in ~8 hours of on-source integration. AIPS software was used for standard calibration and imaging. The resulting uniformly weighted image is thermal noise limited ($1\sigma$ noise level of 8 μJy) and has a restoring beam of 1.00" by 0.69" with a major-axis position angle of $-3.9^{\circ}$. The NE radio jet is shown as contours in Figure 3 overlaid on a color image of the NE x-ray jet, and both jets are seen in projection to be spatially coincident.

---

[1] The National Radio Astronomy Observatory is operated by Associated Universities, Inc., under cooperative agreement with the National Science Foundation.

## 3. DISCUSSION

The R Aqr symbiotic star system is strikingly different at $E < 1$ keV than from 1-7 keV. At the lower energy, the jets dominate. Above 1 keV, R Aqr looks point-like at the Chandra resolution, centered on the star system. The x-ray jets are another manifestation of directed flow seen in the radio, optical, and UV. This is solid evidence for cylindrical accretion geometry, i.e. a disk. Hollis and Koupelis (2000) developed a parcel ejection model for the R Aqr jet emitted along the rotation axis of the massive rotating star-accretion disk system that anchors a magnetic field. A natural extension of the model suggests that the x-ray emission from the jet could be the result of shock emission as the jet parcel plows into the ambient circumstellar medium. The preservation of a well-defined direction for the jets, with evidence for precession, argues against simple spherical accretion. In Figure 1, the strongest parcel in the SW jet seems to have contour compression on its SW edge as would result from SW directed motion. The NE jet component has contour compression on its eastern edge that may be due to counter clockwise motion about the central source reported previously for the NE radio jet (Hollis et al 1997b). In Figure 3, the NE x-ray jet as seen in projection lies along the same trajectory as the NE radio jet, supporting a counter clockwise rotation scenario for the NE x-ray jet. The 430 eV peak in the NE jet and the need for the NEI model to fit the spectrum suggests that this feature was formed recently, with a product of age and ambient density of $4.3 \times 10^{8}$ s cm$^{-3}$. If $n \sim 1$ cm$^{-3}$ as for the typical interstellar medium, the age is ~15 years. This is comparable to the observed timescale for changes in the radio emission, and in x-ray, as inferred from comparing our data with the earlier x-ray satellite observations. For greater density, as might be expected near the R Aqr Mira, the age may be less. The SW x-ray jet is ~30" from the central source compared with ~7.5" for the NE jet, suggesting that it was ejected earlier. It has a different spectrum, with no N peak. We could only fit a model with a very high O abundance. Either there is a velocity shift in the O peak, or there is a different mix of the O VII lines at 0.569, 0.574 and 0.666 keV than our models can reproduce. The derived value of the ionization timescale is about half that of the NE jet, so given the expected variation in ISM density, the time since heating may be about the same. However, if the SW jet was shocked further in the past, we should expect to see relic radio emission as in the NE jet that is not detected. The shock temperatures from the two NEI fits to the jets seem disparate, but only differ by ~1.5$\sigma$, so the difference is not very significant. The central x-ray source has a counterpart in 3.5-cm radio continuum (c.f. Figure 3). The radio emission is a thermal HII region at ~$10^4$ K while the central x-ray source low energy spectrum in Figure 2 fits indicates a temperature of order $10^6$ K. Perhaps the x-ray emitting volume is interior to the HII region, associated with a high energy density region at the hot star/accretion disk. The neutral or lightly ionized Fe at 6.4 keV, without strong associated continuum is evidence for relatively cooler material excited by an unseen source of hard photons or charged particles. Such emission could be reflecting from the surface of the edge-on accretion disk (e.g., Frank, King, & Raine 1992, p. 203), or directly exciting the relatively cool material at the boundary of a surrounding Stromgren surface.

## IV. SUMMARY

We have imaged a two-sided x-ray jet and detected an unresolved central x-ray source in the R Aqr binary system with Chandra ACIS-S3. Non-equilibrium collisional ionization models were applied to the jets. Both jets have ionization timescales of ~$10^8$ s $cm^{-3}$, corresponding to ages of order 10 y or less. A 6.4 keV feature in the central source must be a de-excitation line of relatively cool gas, evidence of an unseen hard source near the hot star, hidden from view owing to the surrounding edge-on accretion disk. Comparison of the NE x-ray jet and central x-ray source with 3.5-cm radio continuum VLA imagery at the same resolution suggests that the x-ray jet may be a high temperature, low density region associated with cooler post-shock UV and radio emitting regions. After more than one hundred years of intense study, R Aqr is still revealing new and fascinating features.

We thank Drs. Eric Feigelson , Steve Drake, John Raymond and Brad Wargelin for useful discussions, and the referee for a thorough reading of the manuscript and suggestions for improvement. We acknowledge support from NASA grant GO0-1062A. E.K. received support from NASA contract NAS8-39073.

## REFERENCES

Anders, E. & Grevesse, N. 1989, Geochimica et Cosmochimica Acta, 53, 197

Arnaud, K.A. 1996, Astronomical Data Analysis Software and Systems V, eds. G.H. Jacoby & J. Barnes, (San Francisco: ASP Conference Series), 101, 17

Borkowski, K. J., Lyerly, W. J.& Reynolds, S. P. 2001, ApJ, 548, 820

CXC (Chandra x-ray Center) 2000, in The Chandra Proposers' Observatory Guide, Version 3.0, TD403.00.003, CXC

Decaux, V., Beiersdorfer, P., Kahn, S. M. & Jacobs, V. L. 1997, ApJ, 482, 1076

Frank, J., King, A., & Raine, D. 1992, Accretion Power in Astrophysics (2 Ed.; Cambridge: Cambridge University Press)

Hollis, J. M., Oliversen, R. J., Michalitsianos, A. G., Kafatos, M., Wagner, R. M. 1991, ApJ, 377, 227

Hollis, J. M., Dorband, J. E. & Yusef-Zadeh, F. 1992, ApJ, 386, 293

Hollis, J.M. & Koupelis, T. 2000 ApJ, 528, 418

Hollis, J.M., Pedelty, J.A., & Kafatos, M. 1997, ApJ, 490, 302

Hollis, J.M., Pedelty, J.A., & Lyon, R.G. 1997, ApJ, 482, L85

Hünsch, M., Schmitt, J. H., Schroeder, K. & Zickgraf, F. 1998, Astron. & Astrophys., 330, 225

Jura, M. & Helfand, D. J. 1984, ApJ, 287, 785

Kaastra, J. S. & Mewe, R. 1993, A&A Suppl., 97, 443

Kafatos, M., Michalitsianos, A. G. & Hollis, J.M. 1986, ApJS, 62, 853

Kellogg , E., Hollis, J. M., Pedelty, J. A. & Lyon, R. G. 2000 American Astronomical Society, HEAD meeting #32, #25.08

Raymond, J.C., & Smith, B.W. 1977, ApJS, 35, 419

Stark, A. A., Gammie, C. F., Wilson, R. W., Bally, J., Linke, R. A., Heiles, C. & Hurwitz, M. 1992, ApJS, 79, 77S

Viotti, R., Piro, L., Friedjung, M., & Cassatella, A. 1987, ApJ, 319, L7

Table 1. Spectral fit parameters[2]

| Jets: Non-equilibrium ionization model with interstellar absorption | | |
|---|---|---|
| Quantity | NE Jet | SW Jet |
| Counts in spectrum, E > 0.3 keV | 396 | 196 |
| C abundance (Fraction of AG) | 1.0 (fixed) | $\leq 0.5$ |
| N abundance (Fraction of AG) | 1.0 (fixed) | 0. (fixed) |
| O abundance (Fraction of AG) | 1.0 (fixed) | 2 0 (fixed) |
| $n_H$ ($10^{20}$ $cm^{-2}$) | $\leq 5.4 \pm 1.5$ | $\leq 10$ |
| kT (keV) | $1.66 \pm 0.99$ | $0.20 \pm 0.06$ |
| T (K) | $19.3 \pm 11.5$ | $2.3 \pm 0.7$ |
| $\tau$, ionization timescale (s $cm^{-3}$) | $4.3 \pm 0.3 \times 10^8$ | $1.9 \pm 0.5 \times 10^8$ |
| Normalization ($10^{-5}$ ph $cm^{-2} s^{-1}$ $keV^{-1}$) | $4.0 \pm 1.0 \times 10^{-5}$ | $1.5(0.8\text{-}7.1) \times 10^{-4}$ |
| Observed flux ($10^{-14}$ ergs $cm^{-2}$ $s^{-1}$) | $7.1 \pm 1.8$ | 2.6(1.4-12.3) |
| Source flux ($n_H = 0$, $10^{-14}$ ergs $cm^{-2}$ $s^{-1}$) | $18.9 \pm 4.7$ | 3.1(1.7-14.7) |
| Reduced $\chi^2$ | 1.04 | 1.19 |
| Confidence (%)[3] | 41 | 31 |

| Center soft source: Blackbody with interstellar absorption | |
|---|---|
| Quantity | Center Source |
| Counts in spectrum, E > 0.3 keV | 108 |
| $n_H$ ($10^{20}$ $cm^{-2}$) | $\leq 6.2$ |
| kT (keV) | $0.18 \pm 0.02$ |
| T ($10^6$K) | $2.09 \pm 0.23$ |
| Normalization ($10^{-5}$ ph $cm^{-2}$ $s^{-1}$ $keV^{-1}$) | $0.009 \pm 0.005$ |
| Observed Flux ($10^{-14}$ ergs $cm^{-2}$ $s^{-1}$, 0.3-2.0 keV) | $0.68 \pm 0.38$ |
| Source Flux, $n_H = 0$ ($10^{-14}$ ergs $cm^{-2}$ $s^{-1}$, 0.3-2.0 keV) | $0.68 \pm 0.38$ |
| Reduced $\chi^2$ | 0.55 |
| Confidence (%) | 84 |

[2] All errors are 1 $\sigma$.

[3] XSPEC null hypothesis probability.

# FIGURE CAPTIONS

Fig. 1. Contour map of the R Aqr 0.1-10 keV x-ray image showing the NE jet, central source, and SW jet. The original ACIS image has been convolved with a 1" FWHM circular gaussian to facilitate smooth contours. The peak photon count in this image is 194.4 counts and the contour levels are 1.4, 2.8, 4.2, 5.6, 8.4, 11.2, 16.8, 22.4, 33.6, 44.8, 89.6, and 179.2 counts. At E>1 keV, the jets are not visible; the dominant emission comes from a point-like source at the star system.

Fig. 2. R Aqr x-ray spectrum extraction regions and spectral fits. The upper panel shows the raw image with source and background spectral extraction regions; *NE2a* and *B1a* for the NE jet, *cntr* and *B2a* for the central source, and *SWa* and *B2a* for the SW jet. The plots show the spectra of the NE jet, central source, and SW jet with best fit models. The NE jet data, binned to 15 counts per channel, has peaks from expected O VII emission at 0.574 keV and N VI at 0.43 keV. The histogram fit is a non-equilibrium ionization model with interstellar absorption. The central source data, binned in 114 eV width channels, is modeled by a black-body, a Fe K$\alpha$ line at ~6.4 keV, K$\beta$ at ~7.0 keV and other discrete but weak lines at K energies of Cr, V, Ca, Ar, and S. The Fe line suggests a hidden hard source associated with an accretion disk. The SW x-ray jet data, binned to 15 counts per channel, has the expected O VII peak at 0.574 keV. The fit is to a non-equilibrium ionization model with high O abundance. (e.g., see Table 4 of Hollis et al 1991).

Fig. 3 - Color rendering of the Chandra ACIS image of the R Aqr NE jet and central HII region overlaid with a contour plot of the 3.5-cm radio continuum emission, observed ten months earlier. The coordinate system is defined relative to the peaks of the x-ray and radio continuum emission (assumed coincident). The x-ray image has been convolved with a 1.00" × 0.69" elliptical gaussian that is the same size as the restoring beam of the radio image. The peak flux is 10.94 mJy beam$^{-1}$ and the contour levels are 0.02, 0.04, 0.06, 0.08, 0.12, 0.16, 0.24, 0.32, 0.48, 0.64, 1.28, 2.56, 5.12, 10.24 mJy beam$^{-1}$.